\documentclass[journal,twoside,web]{ieeecolor}
\usepackage{tmi}
\usepackage{cite}
\usepackage{amsmath,amssymb,amsfonts}
\usepackage{algorithmic}
\usepackage{graphicx}
\usepackage{textcomp}
\usepackage{tabularray} 
\usepackage{ulem} 
\usepackage{booktabs} 
\usepackage{multirow} 
\def\BibTeX{{\rm B\kern-.05em{\sc i\kern-.025em b}\kern-.08em
    T\kern-.1667em\lower.7ex\hbox{E}\kern-.125emX}}
\begin{document}
\title{GLOMIA-Pro: A Generalizable Longitudinal Medical Image Analysis Framework for Disease Progression Prediction}
\author{Shuaitong Zhang, Yuchen Sun, Yong Ao, Xuehuan Zhang, Ruoshui Yang, Jiantao Xu, Zuwu Ai, Haike Zhang, Xiang Yang, Yao Xu, Kunwei Li, Duanduan Chen
\thanks{This work was supported in part by the National Nature Science Foundation of China under Grant 82102140; in part by the Beijing Natural Science Foundation under Grant L232132; and in part by
the Science and Technology Projects of Social Development in Zhuhai under Grant 2320004000149.(Corresponding authors: Duanduan Chen; Kunwei Li.)}
\thanks{Shuaitong Zhang, Yuchen Sun, Xuehuan Zhang, Jiantao Xu, Yao Xu, and
Duanduan Chen are with the School of Medical Technology, Beijing
Institute of Technology, Beijing 100081, China (e-mail:
duanduan@bit.edu.cn).  }
\thanks{Yong Ao is with the Department of Thoracic Surgery, Sun Yat-sen University Cancer Center, Guangzhou, Guangdong 510060, China. }
\thanks{Zuwu Ai, Haike Zhang, Xiang Yang, and
Kunwei Li are with the Department of Radiology, The Fifth Affiliated Hospital of Sun Yat-sen University, Zhuhai, Guangdong 519000, China (e-mail: likunwei@mail.sysu.edu.cn). }
\thanks{Ruoshui Yang is with the Beijing Duanliu Medical Technology Co., Ltd., Beijing 100081, China.}
}

\maketitle

\begin{abstract}
Longitudinal medical images are essential for monitoring disease progression by capturing spatiotemporal changes associated with dynamic biological processes. While current methods have made progress in modeling spatiotemporal patterns, they face three key limitations: (1) lack of generalizable framework applicable to diverse disease progression prediction tasks; (2) frequent overlook of the ordinal nature inherent in disease staging; (3) susceptibility to representation collapse due to structural similarities between adjacent time points, which can obscure subtle but discriminative progression biomarkers. To address these limitations, we propose a \textbf{G}eneralizable \textbf{LO}ngitudinal \textbf{M}edical \textbf{I}mage \textbf{A}nalysis framework for disease Progression prediction (\textbf{GLOMIA-Pro}). GLOMIA-Pro consists of two core components: progression representation extraction and progression-aware fusion. The progression representation extraction module introduces a piecewise orthogonal attention mechanism and employs a novel ordinal progression constraint to disentangle fine-grained temporal imaging variations relevant to disease progression. The progression-aware fusion module incorporates a redesigned skip connection architecture which integrates the learned progression representation with current imaging representation, effectively mitigating representation collapse during cross-temporal fusion. Validated on two distinct clinical applications: knee osteoarthritis severity prediction and esophageal cancer treatment response assessment, GLOMIA-Pro consistently outperforms seven state-of-the-art longitudinal analysis methods. Ablation studies further confirm the contribution of individual components, demonstrating the robustness and generalizability of GLOMIA-Pro across diverse clinical scenarios.
\end{abstract}


\section{Introduction}
\IEEEPARstart{L}{ongitudinal} medical images, where imaging data is collected over multiple time points, are indispensable for evaluating treatment response and tracking disease progression \cite{caruana2015longitudinal}. Compared to cross-sectional medical images, they provide unparalleled insights into complex spatiotemporal patterns associated with dynamic biological processes. In chronic diseases like osteoarthritis, longitudinal images enable early detection of degenerative changes \cite{hu2022adversarial}; while in oncology, they provide noninvasive biomarkers for evalute tumor response to therapies such as neoadjuvant chemoradiotherapy \cite{sun2024lomia}. Therefore, accurate modelling the spatiotemporal patterns in longitudinal medical images is ensential for personalized medicine and individualized treatment decision-making.

Traditional statistical methodologies for longitudinal analysis, including generalized estimating equations \cite{liu2017optimal}, linear mixed-effects models \cite{frings2012quantifying}, and group-based trajectory modeling \cite{nagin2018group}, have primarily emphasized population-level trend characterization. While these approaches effectively model average progression trajectories, they exhibit inherent limitations in generating individualized predictions. Additionally, their dependency on manual feature engineering restricts their capacity to capture complex nonlinear spatiotemporal patterns inherent in high-dimensional longitudinal medical images. In contrast, deep learning has emerged as a transformative paradigm for autonomously extracting intricate spatiotemporal features from longitudinal medical images. Deep learning-based methods for longitudinal medical image analysis are predominantly categorized into two paradigms: \textit{longitudinal co-learning} and \textit{longitudinal fusion learning}. Both methodologies aim to harness spatiotemporal information from longitudinal medical images to predict disease progression, albeit through divergent architectural strategies \cite{baltruvsaitis2018multimodal}.

Longitudinal co-learning methodologies mimic clinical diagnostic workflows, where radiologists assess disease progression through comparative analysis of longitudinal medical images. These approaches explicitly model temporal dependencies among consecutive multiple time-point medical images through two principal strategies: contrastive learning \cite{liang2023modeling,ouyang2022self,emre20243dtinc} and deformable registration techniques \cite{chakravarty2024morph,chen2024longformer}. The former strategy optimize feature spaces where temporally adjacent images exhibit higher similarity while maintaining discriminability across different disease stages. For instance,  \cite{liang2023modeling} leveraged contrastive learning to extract spatiotemporal patterns from longitudinal brain networks, correlating these patterns with cognitive decline in Alzheimer's Disease (AD). The latter technique quantifies voxel-wise intensity variations and spatial deformations. For instance, \cite{chakravarty2024morph} employed deformable registration to enforce anatomical consistency between learned feature displacements and imaging space transformations in Optical Coherence Tomography scans for age-related macular degeneration. Additionally, \cite{yue2022mldrl} proposed multi-loss disentangled representation learning, which disentangled longitudinal radiomics features into shared and unique components to improve esophageal squamous cell carcinoma (ESCC) treatment response prediction. While these methods effectively capture spatiotemporal patterns, they often neglect the ordinal characteristics inherent in disease progression trajectories and exhibit limited generalizability across diverse clinical applications.

Longitudinal fusion learning methodologies emphasize the fusion of temporal dependencies among medical images acquired across distinct timepoints. These approaches employ sequential modeling architectures, including Recurrent Neural Networks (RNN) \cite{xu2019deep}, Temporal Convolutional Networks \cite{konwer2022temporal}, and Transformer \cite{hu2023glim}. \cite{ouyang2020longitudinal} introduced longitudinal pooling to enhance visits representations, thereby improving RNN performance in AD classification. \cite{tong2022dual} and \cite{hu2023glim} integrated spatiotemporal embeddings into Transformer frameworks to model follow-up visit sequences, enabling holistic modeling of disease trajectories. However, longitudinal medical images often exhibit shared semantic attributes, such as anatomical consistency and pathological staging, while disease progression manifests as subtle temporal variations. Conventional sequential architectures frequently fail to resolve these fine-grained progression signals, resulting in \textit{representation collapse} \cite{jing2021understanding}, a phenomenon where features across timepoints converge to degenerate solutions (e.g., identical embeddings $\mathbf{f}_t \approx \mathbf{f}_{t+\Delta t}$), thereby diluting clinically critical progression-related biomarkers. 


In this work, we propose a Generalizable LOngitudinal Medical Image Analysis framework for disease Progression prediction (GLOMIA-Pro). GLOMIA-Pro extracts progression related information from longitudinal images, introduces ordinal progression constraint to align latent representation and mitigates representation collapse through progression-aware fusion. The main contributions of this paper are summarized as follows:
\begin{itemize}
\item This paper proposes a generalizable longitudinal medical image analysis framework for predicting disease progression and validates its effectiveness on two clinically relevant scenarios: knee osteoarthritis severity monitoring and esophageal squamous cell carcinoma treatment response prediction.
\item We introduce piecewise orthogonal attention and a learnable spatiotemporal embbeding to improve disease progression representation from longitudinal medical images; A redesigned skip connection mechanism integrates progression representations with current imaging representation, mitigating representation collapse.
\item We propose a progression constraint that preserves the ordinal nature of disease progression, aligns fine-grained trajectory changes within the broader temporal framework, and enhances the accuracy of disease progression modeling.
\end{itemize}

The proposed framework is evaluated on a public dataset with 7,026 longitudinal knee radiographs for predicting Kellgren and Lawrence grading (KLG) and an in-house data of 208 longitudinal contrast-enhanced CT images for predicting treatment response to neoadjuvant chemoradiotherapy in ESCC. The experimental results demonstrate the proposed method outperforms seven state-of-the-art methods.

This study substantially extends our previous work published in MICCAI \cite{sun2024lomia}, from three main aspects:

1) Methodological Extensions: We fundamentally redesign the optimization framework by introducing an ordinal progression constraint loss function that explicitly encodes ordinal relationships inherent in disease progression patterns. This contrasts with our prior work, which focused solely on binary treatment response-based contrastive loss for ESCC treatment response prediction. Furthermore, we introduce a novel piecewise orthogonal attention to extract the disease progression representation, rather than the cross attention used in \cite{sun2024lomia}. Additionally, we generalize the encoder architecture beyond the original T2T-Transformer by incorporating both CNN and Transformer backbones, systematically validating the framework's adaptability across diverse vision backbones.

2) Rigorous Evaluation: We compare our framework with seven state-of-the-art methods representing different methodological paradigms. Furthermore, we validate the generalizability of our framework on a publicly available knee osteoarthritis dataset, demonstrating robust performance beyond the original ESCC dataset.

3) Expanded Analytical Depth: The experimental analysis is significantly enhanced through (i) ablation studies quantifying the contribution of individual module, and (ii) more detailed analysis and discussion of the experimental results. These extensions provide mechanistic insights into the framework's decision-making processes while addressing clinical interpretability requirements.

\begin{figure*}[t]
    \centerline{\includegraphics{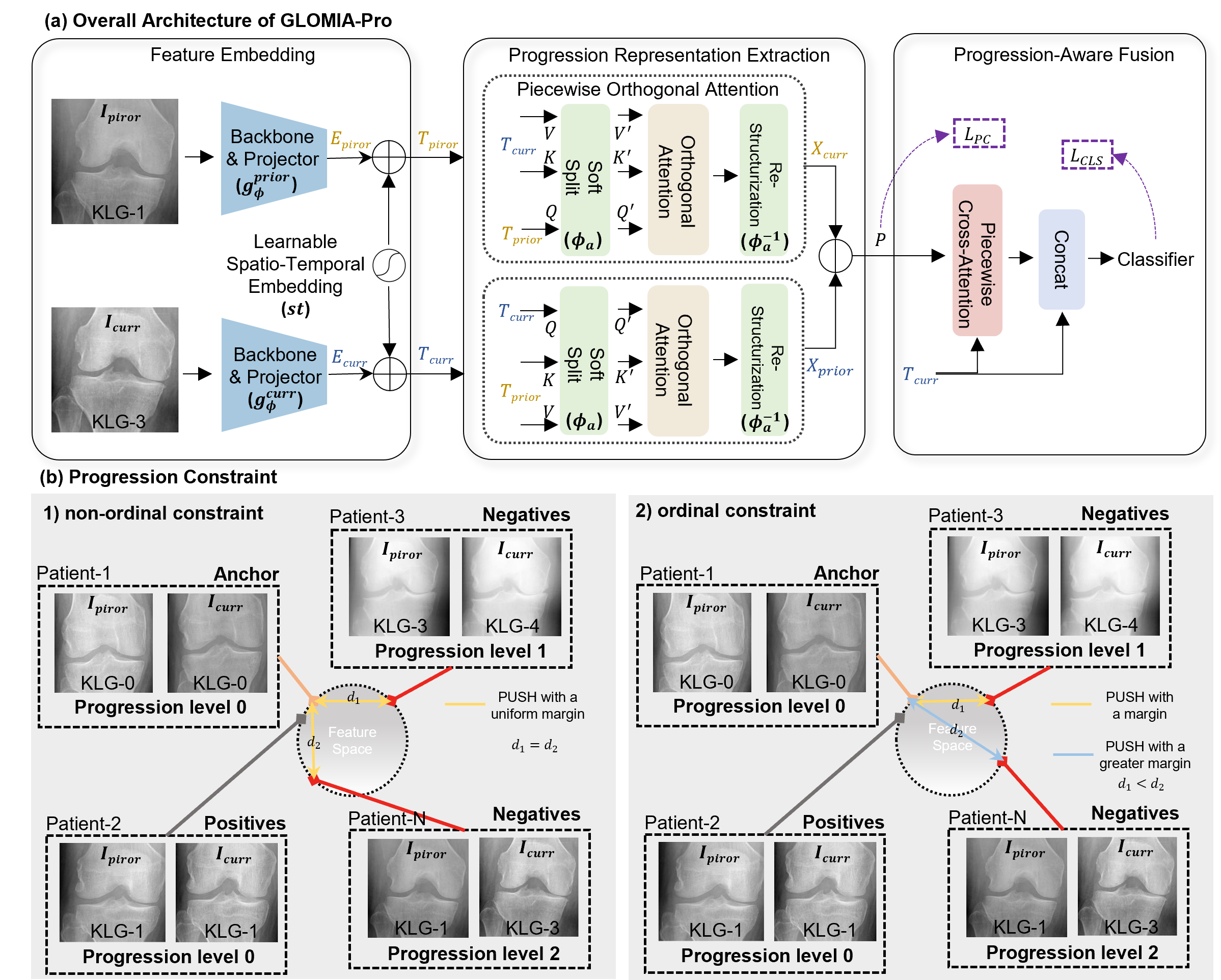}}
    \caption{Illustration of the proposed framework: 
    (a) Overall framework structure. GLOMIA-Pro processes paired images as input and utilizes identical backbone network and projection head architectures for feature embedding. In the progression representation extraction module, the Piecewise Orthogonal Attention (OrA) is employed to extract orthogonal progression information, optimized by ordinal progression constraint. The piecewise mechanism incorporates a soft split operation and its inverse restructurization transformation. Finally, in the progression-aware fusion module, a redesigned skip connection integrates the learned progression representation with the current imaging features to enhance disease progression modeling and mitigate representation collapse.; 
    (b) Compare the Progression Constrain in feature space; (1) in non-ordinal constraint feature space, the representations of different progression levels are pushed with a uniform margin; But (2) in ordinal constraint feature space, the representations of different progression levels are pushed with ordinal ranks. At that situation, the representations of level (0,2) are far away to level (0,1).}
    \label{fig1}
\end{figure*}

\section{METHODS}
The GLOMIA-Pro framework, illustrated in Fig. \ref{fig1}, processes a current visit image \( I_{\text{curr}} \) and a prior visit image \( I_{\text{prior}} \) as paired input. Its architecture comprises three core components: feature embedding for representing each visit image, progression representation extraction module for extracting imaging changes associated with disease progression and progression-aware fusion module for mitigating representation collapse during cross-temporal fusion. The feature embedding module \(g_{\phi}\), comprising a backbone network and a projection head, parameterized by \( \phi \), extracts high-dimensional features and projects them into visual embeddings \( E \in \mathbb{R}^{D} \), formally expressed as:
\begin{equation}
    E_\text{i} = g_{\phi}^{i}(I_{\text{i}}), i \in \{ \text{prior}, \text{curr} \}
\label{eq1}
\end{equation}

The embedding module for \( I_{\text{prior}} \) and \( I_{\text{curr}} \) maintain identical architectures across different vision backbones, with weight-sharing strategies (shared vs. isolated) for parameter \( \phi \) according to dataset-specific characteristics, as quantitatively validated in our ablation studies. The resulting embedding \( E \) are then processed by learnable spatiotemporal transformations.

\subsection{Spatiotemporal Embeddings}
Longitudinal image analysis inherently faces spatiotemporal misalignment challenges due to anatomical variations across different time-point images \cite{liang2023modeling}. Unlike the structured positional dependencies in natural language processing or the sequential ordering of image patches in vision tasks, longitudinal spatiotemporal embeddings do not inherently exhibit explicit
spatial and temporal relationships. While prior works \cite{bannur2023learning,chen2024longformer,tong2022dual} adopted fixed sinusoidal embeddings to encode positional and temporal information, such rigid formulations are often suboptimal for longitudinal medical imaging data where latent representations $\mathbf{E}_{\text{prior}}$ and $\mathbf{E}_{\text{curr}}$ exhibit complex, context-dependent relationships \cite{alayrac2022flamingo}. To address this, we simply use learnable spatiotemporal embeddings (LSTE) $\mathbf{ST} := \{ \mathbf{st}_{\text{prior}};\ \mathbf{st}_{\text{curr}} \} \in \mathbb{R}^{2 \times D}$ to dynamically adapt to inter-visit variations. This parameterization enables alignment of disease progression patterns while preserving spatial coherence. The feature embedding process can be briefly expressed as:
\begin{equation}
    \mathbf{T}_i = \mathbf{E}_i + \mathbf{1}_L \circ \mathbf{st}_i, \quad i \in \{\text{prior},\ \text{curr}\}
\label{eq2}
\end{equation}

Then progression representation module takes \( T_{\text{prior}} \) and \( T_{\text{curr}} \in \mathbb{R}^{D} \) as input to extract disease progression-related information. 

\subsection{Progression Representation Extraction}
Progression representation extraction module mainly consist of Piecewise Orthogonal Attention and an ordinal progression constraint. 

\begin{figure}[!t]
    \centerline{\includegraphics[width=\columnwidth]{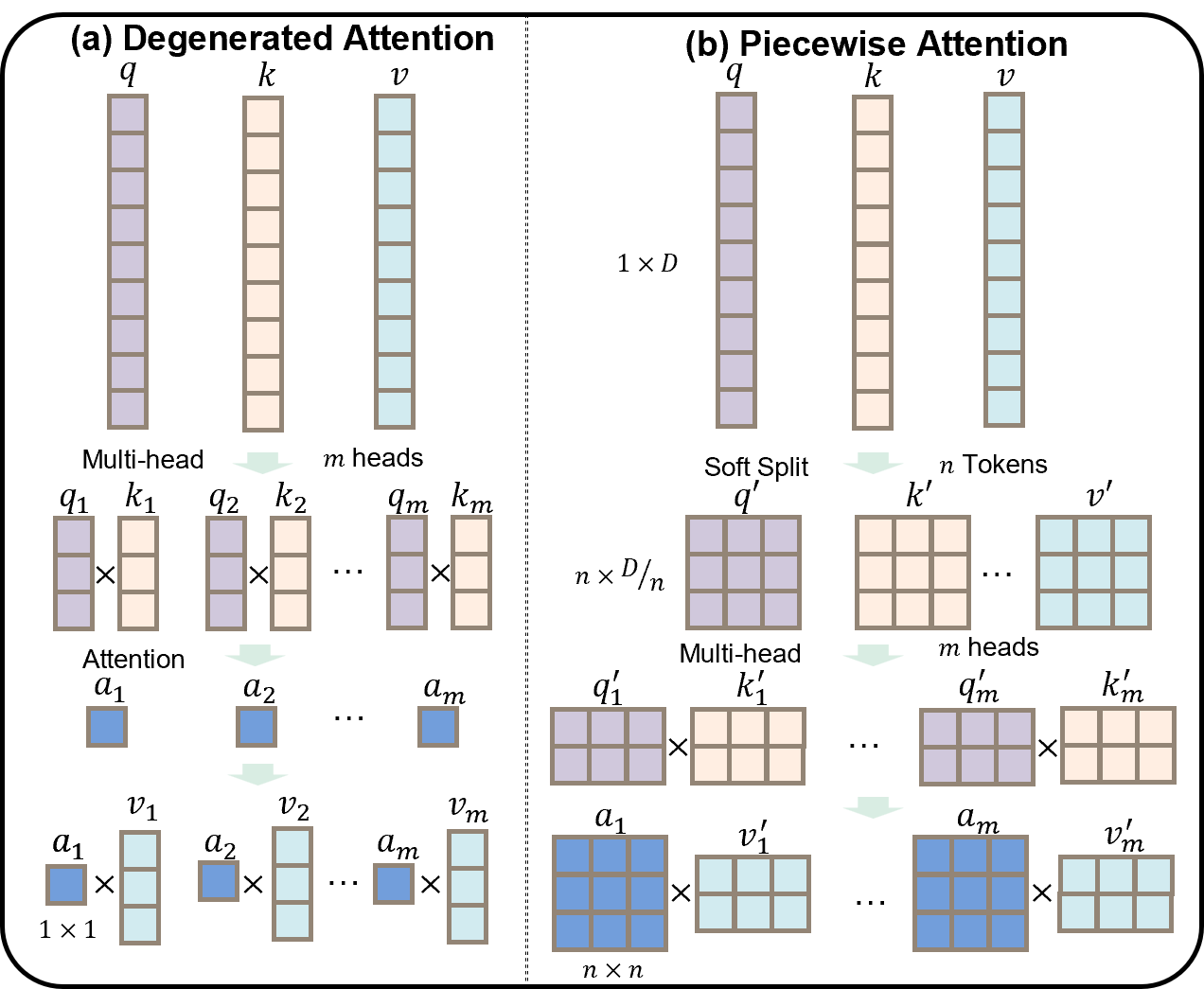}}
    \caption{Schematic illustration of Piecewise Attention and Degenerated Attention. Each column represents a token $\mathbf{qkv} \in \mathbb{R}^{1 \times D}$. 
    \textbf{(a)} Conventional self-attention degeneration with 1D representations: Through multi-head splitting, the queries, keys, and values decompose into $\mathbf{q}_i,\mathbf{k}_i,\mathbf{v}_i \in \mathbb{R}^{1 \times D/m}$ (where $m$ denotes the number of attention heads). The computed attention weights $a \in \mathbb{R}^{1 \times 1}$ degenerated to gated value operations, producing limited-range outputs $[a_1\mathbf{v}_1, a_2\mathbf{v}_2, ..., a_m\mathbf{v}_m]$ that lack long-range feature interactions. 
    \textbf{(b)} Proposed Piecewise Attention mechanism: Decomposes 1D representations into $n$ contextual tokens through Soft Split to gernerate $\mathbf{q'},\mathbf{k'},\mathbf{v'} \in \mathbb{R}^{n \times D/n}$. The computed attention weights $a \in \mathbb{R}^{n \times n}$ enables cross-token attention.
    }
    \label{fig2}
\end{figure}

\subsubsection{Piecewise Orthogonal Attention}
Extracting discriminative progression representation from longitudinal images remains challengings. First, a high level of anatomical redundancy exists between \( T_{\text{prior}} \) and \( T_{\text{curr}}\) \cite{dosovitskiy2020image}. Second, disease-related changes are often localized and subtle, requiring emphasis on divergent features between \( T_{\text{prior}} \) and \( T_{\text{curr}}\). Conventional attention mechanisms often obscure due to anatomical redundancy, which maximize similarity using global cosine similarity $S_C(q, k) = qk^T / \|q\| \|k\|$, where $q,k$ represent tokens from \( T_{\text{prior}} \) and \( T_{\text{curr}}\) during cross-temporal fusion, respectively. To address this, we introduce orthogonal attention (OrA) to emphasize divergent features between them, thereby amplifying disease progression signals. Specifically, we assign higher scores to orthogonal features using $S_O(q, k)=1 - qk^T / \|q\| \|k\|$. OrA operation is defined as:
\begin{equation}
    \text{OrA}(\mathbf{Q}, \mathbf{K}, \mathbf{V}) = \mathbf{J} - \sigma\left(\frac{\mathbf{Q}\mathbf{K}^\top}{\sqrt{d}}\right)\mathbf{V}
    \label{eq:ortho_attn}
\end{equation}
where $\mathbf{J} \in \mathbb{R}^{n \times n}$ is a matrix with all elements equal to 1, $\sigma$ denotes softmax, and $d$ ensures gradient stability. Q, K and V are the queries, keys and values, respectively.

Due to $\mathbf{T}_{\text{prior}}, \mathbf{T}_{\text{curr}} \in \mathbb{R}^D$ are 1D representations, the standard attention mechanism produces attention map which degenerate to scalar gate, as shown in Fig.~\ref{fig2}(a). To overcome this limitation, we introduce a ``piecewise" mechanism that decomposes 1D representations into $n$ contextual tokens, enabling cross-token attention computation. This facilitates the establishment of  global feature dependencies through dynamic feature reassembly. The first step in this process is soft split:
\begin{equation}
    \phi_a: \mathbf{T} \mapsto [\mathbf{T}_1, ..., \mathbf{T}_n] \in \mathbb{R}^{n \times d}, \quad d = D/n
    \label{eq3}
\end{equation}
where $n$ controls interaction granularity. In contrast to vanilla attention that computes scalar gate $\alpha \in \mathbb{R}^{1 \times 1}$ from $\mathbf{q}, \mathbf{k} \in \mathbb{R}^{1 \times D}$, piecewised attention generates attention map $\alpha \in \mathbb{R}^{n \times n}$, establishing long-range dependencies cross temporal features. 

The procedure of the progression representation is illustrated in Fig.~\ref{fig1}. Specifically, \(X_{\text{prior}}^{\text{temp}} \in \mathbb{R}^{n \times d}\) is obtained by applying OrA with \(\phi_a(T_{\text{curr}})\) as Q and \(\phi_a(T_{\text{prior}})\) as K and V, thereby capturing disease progression in \(I_{\text{prior}}\). Similarly, \(X_{\text{curr}}^{\text{temp}}\) is derived by applying orthogonal attention with \(\phi_a(T_{\text{prior}})\) as Q and \(\phi_a(T_{\text{curr}})\) as K and V, capturing disease progression in \(I_{\text{curr}}\). Subsequently, the tokenized progression features $\mathbf{X}_i^{\text{temp}} $ are reconstructed via inverse tokenization ``restructurzation": 

\begin{equation}
    \mathbf{X}_i = \phi_a^{-1}(\mathbf{X}_i^{\text{temp}}), \quad i \in \{\text{prior}, \text{curr}\}
    \label{eq:recon}
\end{equation}
where $\phi_a^{-1}: \mathbb{R}^{n \times d} \rightarrow \mathbb{R}^D$ denotes the inverse tokenization of the piecewised tokenization. The progression representation $\mathbf{P} \in \mathbb{R}^D$ is then defined by element-wise difference, optimized through progression constraint.

\subsubsection{Ordinal Progression Contraint}
The progression constraint $\mathcal{L}_{\text{PC}}$ is designed to align progression representations $\mathbf{P}$ with ordinal progression levels, ensuring that the model captures subtle progression-associated changes. Inspired by \cite{chen2020simple}, we employ a non-linear transformation to convert $\mathbf{P}$ into the latent progression space, where samples from the same progression level are treated as positive pairs, while samples from different levels are treated as negative pairs. In the KLG prediction experiments, progression levels are defined by changes in KLG, while in the ESCC experiments, progression labels correspond to treatment outcomes. Clinically, disease progression is often described as an ordinal classification task, rather than an independent classification task \cite{roy2020deep}. The ordering of progression levels is closely related to the patient's disease severity, confusing two distant grades (e.g., predicting grade 0 to grade 4 in KLG prediction) is more serious than confusing two close grades (e.g., predicting grade 0 to grade 1). Thus, long-distance errors should be penalized more severely than short-distance errors. Moreover, the imaging differences between adjacent progression levels may be subtle, and the network should not be overly penalized for small deviations.

To account for ordinal nature of progression levels, $\mathcal{L}_{\text{PC}}$ builds upon Supcon \cite{khosla2020supervised} by introducing a penalty coefficient \( w \) that adjusts the penalty based on progression level differences. This coefficient assigns a lower penalty to negative pairs with similar progression levels (short-distance levels) and a higher penalty to those negative pairs with larger progression levels. This ensures that the network emphasizes on semantically ordinal similarity and difference. The penalty coefficient \( w_{ib} \) between samples \( i \) and \( b \) is determined by the absolute difference in their progression levels \( l(i) \) and \( l(b) \), with an exponential growth factor governed by the parameter \( \beta \) and a coefficient factor \( \alpha \). Specifically, given a mini-batch of \( N \) input samples, let \( i \in I = \{1, 2, \ldots, N\} \) be the index of any sample in the mini-batch, which is treated as the anchor. Let \( b \in B(i) \equiv I \setminus \{i\} \) represent the set of all other indices. The penalty coefficient can be defined as:
\begin{equation}
    \begin{aligned}
        w_{ib} = 
        \begin{cases}
        1, & \text{if } l(i) = l(b) \\
        a \cdot \beta^{|l(i)-l(b)|}, & \text{if } l(i) \neq l(b)
        \end{cases}
    \end{aligned}
\end{equation}

The final similarity is computed as the product of the $w$ and the original similarity in \cite{khosla2020supervised}. Given a progression \( p_i \), the similarity between \( p_i \) and its positive sample \( p_j \) is summed and divide by the similarity between \( p_i \) and all other \( N-1 \) samples, including both positive pairs and negative pairs. To avoid division by zero, we ensure that the number of samples with same progression level is \( \ge 2 \) in mini-batch during sampling. The loss can be defined as:
\begin{equation}
    \ell(i,j)=-\log\frac{\exp\left(\operatorname{S}\left(p_i,p_j\right)/\tau\right)}{\sum_{b=1}^{N} \exp\left(w_{ib}\operatorname{S}\left(p_i,p_b\right)/\tau\right)}
\end{equation}
where \( j \in J(i) \) be the positive sample of \( i \), which has the same progression level. \( \text{S}(u, v) = u^T v / \|u\| \|v\| \) denotes cosine similarity between \( u \) and \( v \), and \( \tau \) is a temperature parameter. Overall, the supervised ordianl progression contraint is expressed as:
\begin{equation}
    \mathcal{L}_{PC}=\sum_{i\in I}\frac{1}{|J(i)|}\sum_{j\in J(i)}\ell(i,j)
\end{equation}

\subsection{Progressive-aware Fusion}
Building upon our previous findings regarding temporal importance attribution \cite{sun2024lomia}, we propose a progression-aware fusion mechanism that employs skip connections to combine multi-scale features while preventing representation collapse. In this design, $\mathbf{P}$ serves as query, while $\mathbf{T}_{\text{curr}}$ serves as both key and value in the Piecewise Cross-Attention module. The module adopts the same tokenization and reconstruction procedures ($\phi_a$ and $\phi_a^{-1}$) as used in our Piecewise Orthogonal Attention, but replaces orthogonal attention with cross-attention. To preserve recent temporal information, we concatenate the cross-attention output with $\mathbf{T}_{\text{curr}}$ before final classification. 

For the final output of the framework, KLG or tumor treatment response is optimized by the classification loss \( \mathcal{L}_{\text{CLS}} \). In the KOA experiment, the ordinal classification loss is chosen \cite{chen2019fully}. In the ESCC experiment, the focal loss is selected \cite{lin2017focal}, consistent with our previous work \cite{sun2024lomia}. The overall loss is a weighted sum of ordinal progression contraint \( \mathcal{L}_{\text{PC}} \) and classification loss \( \mathcal{L}_{\text{CLS}} \):
\begin{equation}
    \mathcal{L}=\lambda_1 \mathcal{L}_{\text{PC}}+\lambda_2 \mathcal{L}_{\text{CLS}}
\end{equation}

\begin{table}[t] 
    \centering
    \caption{Data distribution of KOA and ESCC datasets.}
    \begin{tabular}{llll} 
        \toprule
        \multicolumn{1}{c}{}                                                                       & \multicolumn{3}{c}{KOA}                            \\
        \multirow{3}{*}{\begin{tabular}[c]{@{}l@{}}Basic\\Information\end{tabular}}                & \#. of patients~ ~ ~~  & 3513       &              \\
                                                                                                   & \#. of knees           & 7026       &              \\
                                                                                                   & \#. of images          & 14052      &              \\ 
        \midrule
        \multirow{5}{*}{Baseline}                                                                  & KLG - 0                & 2800       & (39.6\%)     \\
                                                                                                   & KLG - 1                & 1266       & (45.2\%)     \\
                                                                                                   & KLG - 2                & 1803~      & (25.7\%)     \\
                                                                                                   & KLG - 3                & 937~~      & (13.3\%)     \\
                                                                                                   & KLG - 4                & 220~ ~     & ( 3.1\%)     \\ 
        \midrule
        \multirow{5}{*}{12-month}                                                                  & KLG - 0                & 2694       & (38.3\%)     \\
                                                                                                   & KLG - 1                & 1209       & (17.2\%)     \\
                                                                                                   & KLG - 2                & 1822       & (25.9\%)     \\
                                                                                                   & KLG - 3                & 1008       & (14.3\%)  \\
                                                                                                   & KLG - 4                & 293~       & ( 4.2\%)     \\ 
        \midrule
        \multicolumn{1}{c}{}                                                                       & \multicolumn{3}{c}{ESCC}                           \\
        \multirow{4}{*}{\begin{tabular}[c]{@{}l@{}}Basic\\Information\end{tabular}}                & \#. of patients ~ ~    & 208        &              \\
                                                                                                   & \#. of images          & 519        &              \\
                                                                                                   & Female/Male            & 178/30     &              \\
                                                                                                   & Age(mean±std)          & 59.6 ± 6.3 &              \\ 
        \midrule
        \multirow{5}{*}{\begin{tabular}[c]{@{}l@{}}Distribution\\and Tumor\\Location\end{tabular}} & pCR                    & 95         & (45.7\%)     \\
                                                                                                   & non-pCR                & 113        & (54.3\%)     \\
                                                                                                   & Upper thoracic region  & 35         & (16.8\%)     \\
                                                                                                   & Middle thoracic region & 106        & (51.0\%)     \\
                                                                                                   & Lower thoracic region  & 67         & (32.2\%)     \\
        \bottomrule
        \end{tabular}
    \label{data-distribution}
\end{table}

\section{EXPERIMENTS AND RESULTS}

\begin{table*}[t]\scriptsize 
    \centering
    \caption{COMPARISON OF PERFORMANCE [95\% CI] WITH DIFFERENT LONGITUDINAL METHODS ON KOA AND ESCC DATASETS.}
    \begin{tabular}{lccccccc} 
    \toprule
    \multirow{2}{*}{Methods} & \multicolumn{4}{c}{KOA} & \multicolumn{3}{c}{ESCC} \\
    \cmidrule(r){2-5} \cmidrule(l){6-8}
    & BAC (\%) & ACC (\%) & Kappa & MSE & AUC (\%) & ACC (\%) & MCC \\ 
    \midrule
    LP\cite{ouyang2020longitudinal}                      & \begin{tabular}[c]{@{}c@{}}75.78\\{[}73.03 - 78.54]\end{tabular}  & \begin{tabular}[c]{@{}c@{}}75.76\\{[}73.52 - 78.00]\end{tabular} & \begin{tabular}[c]{@{}c@{}}0.8970\\{[}0.8827 - 0.9111]\end{tabular} & \begin{tabular}[c]{@{}c@{}}0.3005\\{[}0.2637 - 0.3372]\end{tabular} & \begin{tabular}[c]{@{}c@{}}83.83\\{[}81.11 - 86.55]\end{tabular} & \begin{tabular}[c]{@{}c@{}}74.04\\{[}71.01 - 77.09]\end{tabular} & \begin{tabular}[c]{@{}c@{}}0.4784\\{[}0.4185 - 0.5390]\end{tabular}  \\
    DiT\cite{tong2022dual}                      & \begin{tabular}[c]{@{}c@{}}73.22\\{[}70.49 - 75.90]\end{tabular}  & \begin{tabular}[c]{@{}c@{}}72.29\\{[}69.91 - 74.62]\end{tabular} & \begin{tabular}[c]{@{}c@{}}0.8805\\{[}0.8652 - 0.8953]\end{tabular} & \begin{tabular}[c]{@{}c@{}}0.3558\\{[}0.3176 - 0.3945]\end{tabular} & \begin{tabular}[c]{@{}c@{}}83.31\\{[}80.50 - 86.05]\end{tabular} & \begin{tabular}[c]{@{}c@{}}76.44\\{[}73.47 - 79.36]\end{tabular} & \begin{tabular}[c]{@{}c@{}}0.5257\\{[}0.4656 - 0.5844]\end{tabular}  \\
    LOMIA-T\cite{sun2024lomia}                 & \begin{tabular}[c]{@{}c@{}}73.59\\{[}71.06 - 76.11]\end{tabular}  & \begin{tabular}[c]{@{}c@{}}75.05\\{[}72.77 - 77.32]\end{tabular} & \begin{tabular}[c]{@{}c@{}}0.8462\\{[}0.8250 - 0.8670]\end{tabular} & \begin{tabular}[c]{@{}c@{}}0.4869\\{[}0.4272 - 0.5463]\end{tabular} & \begin{tabular}[c]{@{}c@{}}86.85\\{[}84.37 - 89.48]\end{tabular} & \begin{tabular}[c]{@{}c@{}}75.96\\{[}73.07 - 79.16]\end{tabular} & \begin{tabular}[c]{@{}c@{}}0.5224\\{[}0.4653 - 0.5852]\end{tabular}  \\
    LongFormer\cite{chen2024longformer} & \begin{tabular}[c]{@{}c@{}}76.44\\{[}74.12 - 78.76]\end{tabular}  & \begin{tabular}[c]{@{}c@{}}76.72\\{[}74.02 - 79.42]\end{tabular} & \begin{tabular}[c]{@{}c@{}}0.8912\\{[}0.8765 - 0.9059]\end{tabular} & \begin{tabular}[c]{@{}c@{}}0.2960\\{[}0.2597 - 0.3323]\end{tabular} & \begin{tabular}[c]{@{}c@{}}85.74\\{[}82.77 - 88.71]\end{tabular}   & \begin{tabular}[c]{@{}c@{}}76.44\\{[}73.67 - 79.21]\end{tabular} & \begin{tabular}[c]{@{}c@{}}0.5298\\{[}0.4634 - 0.5962]\end{tabular}  \\
    TLSTM\cite{yin2022predicting}  & \begin{tabular}[c]{@{}c@{}}74.62\\{[}71.81 - 77.43]\end{tabular}  & \begin{tabular}[c]{@{}c@{}}75.12\\{[}72.85 - 77.37]\end{tabular} & \begin{tabular}[c]{@{}c@{}}0.8941\\{[}0.8798 - 0.9079]\end{tabular} & \begin{tabular}[c]{@{}c@{}}0.3147\\{[}0.2784 - 0.3511]\end{tabular} & -                                                                & -                                                                & -                                                                 \\
    RP-Net\cite{rosenberg2018prediction} & -                                                                 & -                                                                & -                                                                & -                                                                & \begin{tabular}[c]{@{}c@{}}79.87\\{[}76.79 - 82.96]\end{tabular} & \begin{tabular}[c]{@{}c@{}}75.00\\{[}72.01 - 78.00]\end{tabular} & \begin{tabular}[c]{@{}c@{}}0.4962\\{[}0.4361 - 0.5566]\end{tabular}  \\
    MLDRL\cite{yue2022mldrl}  & -                                                                 & -                                                                & -                                                                & -                                                                &  \begin{tabular}[c]{@{}c@{}}86.55\\{[}83.54 - 89.56]\end{tabular} & \begin{tabular}[c]{@{}c@{}}\textbf{80.96}\\{[}76.74 - 85.19]\end{tabular} & -  \\ 
    \midrule
    \begin{tabular}[l]{@{}l@{}}GLOMIA-Pro\end{tabular}                    & \begin{tabular}[c]{@{}c@{}}\textbf{78.63}\\{[}76.23 - 81.00]\end{tabular} & \begin{tabular}[c]{@{}c@{}}\textbf{76.97}\\{[}74.77 - 79.14]\end{tabular} & \begin{tabular}[c]{@{}c@{}}\textbf{0.9045}\\{[}0.8913 - 0.9173]\end{tabular} & \begin{tabular}[c]{@{}c@{}}\textbf{0.2856}\\{[}0.2518 - 0.3199]\end{tabular} & \begin{tabular}[c]{@{}c@{}}\textbf{87.90}\\{[}85.61 - 90.18]\end{tabular} & \begin{tabular}[c]{@{}c@{}}78.85\\{[}76.00 - 81.70]\end{tabular} & \begin{tabular}[c]{@{}c@{}}\textbf{0.5816}\\{[}0.5265 - 0.6365]\end{tabular}  \\
    \bottomrule
    \end{tabular}
    \label{tab1}
\end{table*}

\begin{figure*}[t]
    \centerline{\includegraphics[width=\textwidth]{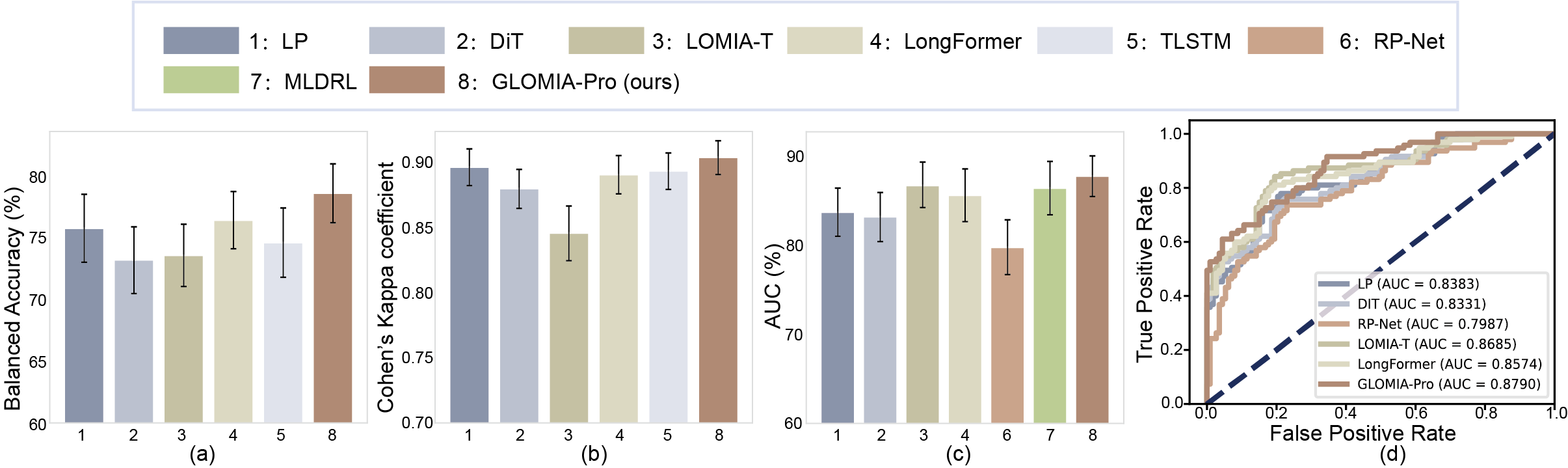}}
    \caption{Performance of different longitudinal methods on KOA and ESCC datasets. (a) BAC of different methods on KOA dataset; (b) Kappa of different methods on KOA dataset; (c) AUC of different methods on ESCC dataset; (d) ROC curve of different methods on ESCC dataset.
    }
    \label{fig3}
\end{figure*}

\begin{figure}[t]
    \centering
    \centerline{\includegraphics[width=\columnwidth]{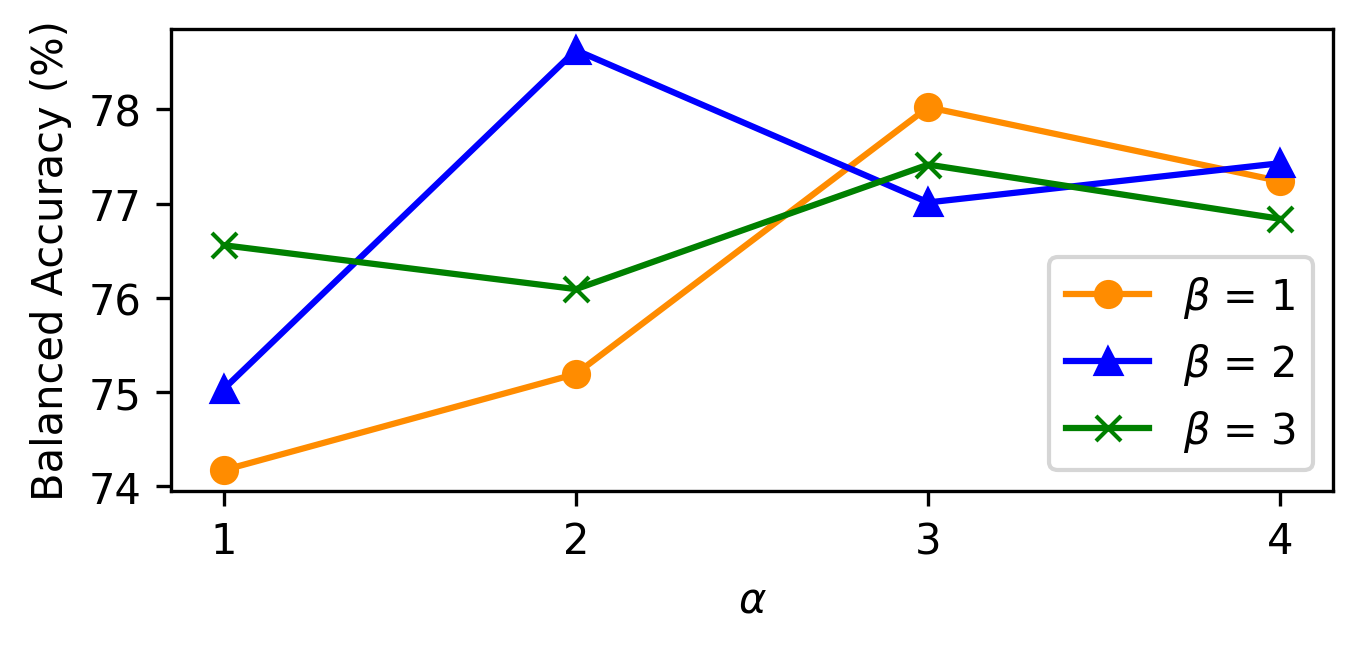}}
    \caption{Process of hyperparameters search for $\mathcal{L}_{\text{PC}}$}
    \label{hyper-param}
\end{figure}

\begin{table}[t]
    \centering
    \caption{PERFORMANCE ON EARLY OA DIAGNOSIS RESULTS.}
    \begin{tabular}{lcc} 
    \toprule
    Methods      & ACC (\%)                     & F1 (\%)                       \\
    \midrule
    DRAE\cite{nasser2020discriminative} & 82.53 \par{}{[}81.66 - 83.40] & 83.48 \par{}{[}82.62 - 84.34]  \\
    GLOMIA-Pro        & \textbf{87.32} \par{}{[}85.17 - 89.50] & \textbf{85.92} \par{}{[}83.37 - 88.49]  \\
    \bottomrule
    \end{tabular}
    \label{tab2}
\end{table}

\subsection{Datasets}
To validate our proposed framework, we perform experiments on two clinically relevant scenarios: (1) predicting the severity of knee osteoarthritis (KOA) from 2D longitudinal knee radiographs, and (2) predicting the pathological complete response (pCR) in esophageal squamous cell carcinoma (ESCC) patients received neoadjuvant chemoradiotherapy (nCRT) using 3D longitudinal contrast-enhanced CT images.

KOA represents a global healthcare crisis, as the leading cause of chronic disability among adults over 60 years \cite{heidari2011knee}. The KLG system is employed for severity assessment in clinical practice, classifying individual joints into one of five grades, with 0 representing normal and 4 being the most severe \cite{kellgren1957radiological}. Accurate diagnosis and prediction of KLG are essential for early intervention, which can significantly improving patients' quality of life \cite{cross2014global}.

ESCC constitutes approximately 90\% of esophageal cancer cases in China\cite{bray2018global}, with nCRT followed by esophagectomy being the standard treatment option for locally advanced ESCC\cite{van2012preoperative}. Evidences demonstrate that a wait-and-see strategy is more appropriate for patients who achieve pCR after nCRT, rather than esophagectomy \cite{van2013recurrence}. Therefore, preoperative prediction of treatment response (pCR or non-pCR) in individual ESCC patients is of clinical significance for individualized treatment decison-making.

\subsubsection{KOA Dataset}
We utilize a multi-center longitudinal dataset from the Osteoarthritis Initiative (OAI) \cite{nevitt2006osteoarthritis}, comprising knee radiographs from 4,796 participants. Each radiograph is annotated with KLG by a physician interpreter. After excluding cases with missing multi-time point scans in baseline and 12-month scans, our study includes 3513 participants (7026 longitudinal knee radiographs). These radiographs are divided randomly into train, validation, and test sets at a ratio of 7:1:2. Knee joint landmarks are located by KneeLocalizer tool \cite{tiulpin2017novel}, with manual refinement to ensure landmark accuracy. All images are resized to a resolution of \(1024\times1024\), and the input image size for the model is set to \(896\times896\), which is consistent with \cite{zhang2020attention}. 

\begin{table*}[!t]
    \centering
    \caption{VALIDATE THE GENERALIZABILITY OF PROPOSED FRAMEWORK WITH DIFFERENT VISION BACKBONES ON KOA DATASETS.}
    \begin{tabular}{lcccc} 
        \toprule
        Methods            & BAC (\%)                        & ACC   (\%)                       & Kappa                      & MSE                       \\ 
        \midrule
        ResCBAM\cite{zhang2020attention}            & 74.39 [71.75 - 77.02] & 74.42 [72.15 - 76.66]  & 0.8797 [0.8637 - 0.8952] & 0.3685 [0.3263 - 0.4113]  \\
        Proposed framework + \cite{zhang2020attention}    & \textbf{78.30} [75.99 - 80.59] & \textbf{75.97} [73.72 - 78.21]  & \textbf{0.9051} [0.8924 - 0.9174] & \textbf{0.2849} [0.2528 - 0.3170]  \\
        \midrule
        Densenet169\cite{thomas2020automated}        & 71.74 [68.91 - 74.53] & 73.64 [71.31 - 75.91]  & 0.8764 [0.8598 - 0.8925] & 0.3777 [0.3335 - 0.4225]  \\
        Proposed framework + \cite{thomas2020automated} & \textbf{76.49} [74.07 - 78.90] & \textbf{73.35} [71.04 - 75.66]  & \textbf{0.8902} [0.8761 - 0.9039] & \textbf{0.3260} [0.2909 - 0.3613]  \\
        \midrule
        Swin-Transformer\cite{liu2021swin}               & 70.23 [67.41 - 73.05] & 72.93 [70.61 - 75.12]  & 0.8661 [0.8482 - 0.8832] & 0.4146 [0.3673 - 0.4627]  \\
        Proposed framework + \cite{liu2021swin}     & \textbf{74.07} [71.29 - 76.82] & \textbf{74.77} [72.49 - 77.02]  & \textbf{0.8860} [0.8706 - 0.9008] & \textbf{0.3459} [0.3059 - 0.3862]  \\
        \midrule
        ResNet50 (Baseline) \cite{he2016deep}           & 74.27 [71.61 - 76.95] & 73.99 [71.72 - 76.26]  & 0.8816 [0.8658 - 0.8970] & 0.3614 [0.3192 - 0.4036]  \\ 
        Proposed framework + \cite{he2016deep}               & \textbf{78.63} [76.23 - 81.00] & \textbf{76.97} [74.77 - 79.14] & \textbf{0.9045} [0.8913 - 0.9173] & \textbf{0.2856} [0.2518 - 0.3199]  \\
        \bottomrule
        \end{tabular}
    \label{tab3}
\end{table*}

\begin{table}[t] \scriptsize
    \centering
    \caption{VALIDATE THE GENERALIZABILITY OF PROPOSED FRAMEWORK WITH DIFFERENT VISION BACKBONES ON ESCC DATASETS.}
    \begin{tabular}{lccc} 
        \toprule
        Methods   & AUC (\%)                                                                & ACC (\%)                                                                & MCC                                                                  \\ 
        \midrule
        T2T \cite{yuan2021tokens}      & \begin{tabular}[c]{@{}c@{}}70.64\\{[}67.05 - 74.18]\end{tabular} & \begin{tabular}[c]{@{}c@{}}66.83\\{[}63.57 - 70.06]\end{tabular} & \begin{tabular}[c]{@{}c@{}}0.3266\\{[}0.2608 - 0.3920]\end{tabular}  \\ 
        \begin{tabular}[l]{@{}l@{}}Proposed\\framework+\cite{yuan2021tokens}\end{tabular} & \begin{tabular}[c]{@{}c@{}}80.70\\{[}77.70 - 83.65]\end{tabular} & \begin{tabular}[c]{@{}c@{}}73.56\\{[}70.48 - 76.61]\end{tabular} & \begin{tabular}[c]{@{}c@{}}0.4676\\{[}0.4058 - 0.5292]\end{tabular}  \\
        \midrule
        \begin{tabular}[l]{@{}l@{}}ResNet50\\(Baseline) \cite{he2016deep}\end{tabular}  & \begin{tabular}[c]{@{}c@{}}72.70\\{[}69.25 - 76.13]\end{tabular} & \begin{tabular}[c]{@{}c@{}}66.35\\{[}63.07 - 69.63]\end{tabular} & \begin{tabular}[c]{@{}c@{}}0.3470\\{[}0.2832 - 0.4110]\end{tabular}  \\ 
        \begin{tabular}[l]{@{}l@{}}Proposed\\framework+\cite{he2016deep}\end{tabular}     & \begin{tabular}[c]{@{}c@{}}\textbf{87.90}\\{[}85.61 - 90.18]\end{tabular} & \begin{tabular}[c]{@{}c@{}}\textbf{78.85}\\{[}76.00 - 81.70]\end{tabular} & \begin{tabular}[c]{@{}c@{}}\textbf{0.5816}\\{[}0.5265 - 0.6365]\end{tabular}  \\
        \bottomrule
        \end{tabular}
\label{tab4}
\end{table}

\subsubsection{ESCC Dataset}
This dataset includes 209 locally advanced ESCC patients from two different hospitals: The Fifth Affiliated Hospital of Sun Yat sen University and Cancer center of Sun Yat sen University. All ESCC patients underwent contrast-enhanced CT examinations before and after nCRT. Tumor regions on both timepoint CT images are delineated using the method described in \cite{zhang2024deep}, with all delineations verified and corrected by an twenty-year-experience  radiologists (K.W. Li). All tumor regions are resampled and padded to achieve uniform dimensions of \(32\times64\times64\). This study is in accordance with the Declaration of Helsinki and is approved by the ethics committee of the enrolled hospital.

\begin{figure}[t]
    \centerline{\includegraphics[width=\columnwidth]{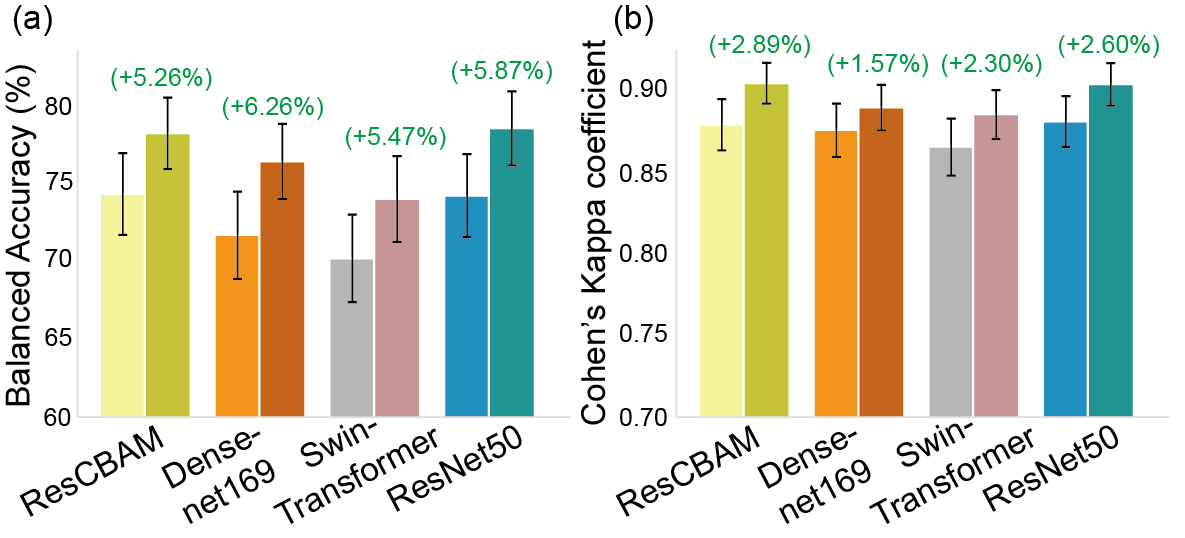}}
    \caption{Performance Improvements of the Proposed Framework For Applying Different Vision Backbones on KOA Datasets. (a) BAC of different backbone on KOA dataset; (b) Kappa of different backbone on KOA dataset. Left: cross-section backbone; Right: proposed framework + backbone.  
    }
    \label{fig5}
\end{figure}

\begin{figure}[t]
    \centerline{\includegraphics{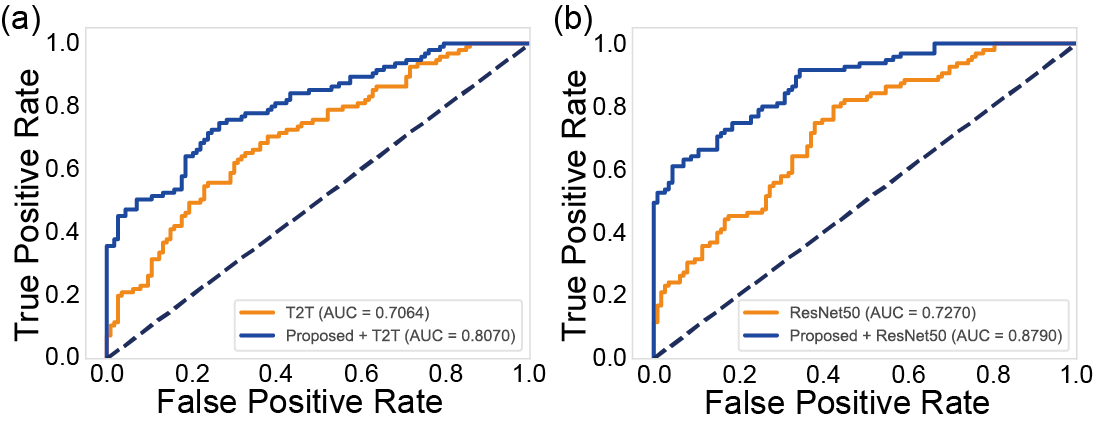}}
    \caption{AUCs Improvements of the Proposed Framework For Applying Different Vision Backbones on ESCC Datasets. (a) ROC curve of T2T-Transfomer as backbone; (b) ROC curve of ResNet50 as backbone.
    }
    \label{fig6}
\end{figure}

\subsection{Experimental Settings}
We employ AdamW optimizer ($\beta_1=0.9$, $\beta_2=0.999$, weight decay = 0.03) with a cosine learning rate scheduler over 50 epochs. A balanced loss combination is utilized, with $\lambda_1=\lambda_2=0.5$ for both tasks. For KOA assessment, we initialize the model with ImageNet-pretrained weights from TorchVision 0.19.1, training with an initial learning rate of $5\times10^{-6}$ and a batch size of 16. In contrast, ESCC evaluation uses a learning rate of $3\times10^{-4}$ and a batch size of 64. The weights of backbone network are pretrained using an auxiliary discrimination task \cite{he2020momentum} to mitigate data scarcity of ESCC data. The ordinal progression constraint loss $\mathcal{L}_{\text{PC}}$ employs hyperparameters \( \alpha = 2 \) and \( \beta = 2 \) for KOA (illustrated in Fig.~\ref{hyper-param}). All experiments use a fixed temperature parameter \( \tau = 0.07 \), 16 tokens in the piecewise mechanism, and a 0.4 dropout rate.  Implementation is conducted using PyTorch 2.4.1 \cite{paszke2019pytorch}, leveraging two NVIDIA GTX 4090 (KOA dataset) and two A100 (ESCC dataset) GPUs. The code will be publicly available at https://github.com//syc19074115//GLOMIA.  

Comprehensive evaluations are performed on two clinical datasets to assess the model's effectiveness in predicting KLG for KOA and pCR for ESCC. For the ESCC dataset, five-fold cross-validation is applied, reporting classification accuracy (ACC), Matthews correlation coefficient (MCC), and area under the receiver operating characteristic curve (AUC) against state-of-the-art methods. The KOA evaluation uses a held-out test set with balanced accuracy (BAC), ACC, mean squared error (MSE), and Cohen's kappa coefficient (Kappa). To ensure statistical reliability, all performance metrics are presented with 95\% confidence intervals (CI) derived from 10,000 bootstrap resamples.

\begin{table*}[h] 
    \centering
    \caption{ABALATION STUDY ON THE EFFECTIVENESS OF THE ISOLATION FOR BACKBONES, SPATIOTEMPORAL EMBEDDINGS AND THE PROPOSED KEY MODULES}
    \begin{tabular}{lccccccc}
        \toprule
    \multirow{2}{*}{Methods} & \multicolumn{4}{c}{KOA} & \multicolumn{3}{c}{ESCC} \\
    \cmidrule(r){2-5} \cmidrule(l){6-8}
    & BAC & ACC & Kappa & MSE & AUC & ACC & MCC \\ 
    \midrule
        \begin{tabular}[l]{@{}l@{}}Embedding\\(isolated)\end{tabular} & \begin{tabular}[c]{@{}c@{}}75.92\\{[}73.40 - 78.42]\end{tabular} & \begin{tabular}[c]{@{}c@{}}73.56\\{[}71.26 - 75.85]\end{tabular}  & \begin{tabular}[c]{@{}c@{}}0.8899\\{[}0.8753 - 0.9039]\end{tabular} & \begin{tabular}[c]{@{}c@{}}0.3296\\{[}0.2923 - 0.3673]\end{tabular} & \begin{tabular}[c]{@{}c@{}}\textbf{87.90$^1$}\\{[}85.61 - 90.18]\end{tabular} & \begin{tabular}[c]{@{}c@{}}\textbf{78.85$^1$}\\{[}76.00 - 81.70]\end{tabular} & \begin{tabular}[c]{@{}c@{}}\textbf{0.5816$^1$}\\{[}0.5265 - 0.6365]\end{tabular}  \\
        \begin{tabular}[l]{@{}l@{}}Embedding  \ \ \ \ \  \\(shared)\end{tabular}  & \begin{tabular}[c]{@{}c@{}}\textbf{78.63$^2$}\\{[}76.23 - 81.00]\end{tabular} & \begin{tabular}[c]{@{}c@{}}\textbf{76.97$^2$}\\{[}74.77 - 79.14]~\end{tabular} & \begin{tabular}[c]{@{}c@{}}\textbf{0.9045$^2$}\\{[}0.8913 - 0.9173]\end{tabular} & \begin{tabular}[c]{@{}c@{}}\textbf{0.2856$^2$}\\{[}0.2518 - 0.3199]\end{tabular} & \begin{tabular}[c]{@{}c@{}}84.17\\{[}81.57 - 86.99]\end{tabular} & \begin{tabular}[c]{@{}c@{}}73.08\\{[}70.13 - 76.43]\end{tabular} & \begin{tabular}[c]{@{}c@{}}0.4646\\{[}0.4068 - 0.5305]\end{tabular}  \\
        \midrule
        TPE\cite{hu2023glim}  & \begin{tabular}[c]{@{}c@{}}77.69\\{[}75.18 - 80.17]\end{tabular} & \begin{tabular}[c]{@{}c@{}}76.68\\{[}74.47 - 78.87]\end{tabular}  & \begin{tabular}[c]{@{}c@{}}0.9029\\{[}0.8892 - 0.9162]\end{tabular} & \begin{tabular}[c]{@{}c@{}}0.2920\\{[}0.2562 - 0.3280]\end{tabular} & \begin{tabular}[c]{@{}c@{}}81.51\\{[}78.41 - 84.30]\end{tabular} & \begin{tabular}[c]{@{}c@{}}75.48\\{[}72.40 - 78.46]\end{tabular} & \begin{tabular}[c]{@{}c@{}}0.5042\\{[}0.4417 - 0.5648]\end{tabular}  \\
        TDPE\cite{tong2022dual} & \begin{tabular}[c]{@{}c@{}}76.62\\{[}73.97 - 79.28]\end{tabular} & \begin{tabular}[c]{@{}c@{}}74.27\\{[}72.00 - 76.56]\end{tabular}  & \begin{tabular}[c]{@{}c@{}}0.8920\\{[}0.8779 - 0.9059]\end{tabular} & \begin{tabular}[c]{@{}c@{}}0.3097\\{[}0.2744 - 0.3447]\end{tabular} & \begin{tabular}[c]{@{}c@{}}83.08\\{[}80.35 - 86.01]\end{tabular} & \begin{tabular}[c]{@{}c@{}}76.44\\{[}73.63 - 79.60]\end{tabular} & \begin{tabular}[c]{@{}c@{}}0.5237\\{[}0.4670 - 0.5873]\end{tabular}  \\
        TIPE\cite{sun2024lomia} & \begin{tabular}[c]{@{}c@{}}77.48\\{[}75.13 - 79.80]\end{tabular} & \begin{tabular}[c]{@{}c@{}}73.99\\{[}71.68 - 76.29]\end{tabular}  & \begin{tabular}[c]{@{}c@{}}0.8936\\{[}0.8796 - 0.9072]\end{tabular} & \begin{tabular}[c]{@{}c@{}}0.3210\\{[}0.2844 - 0.3578]\end{tabular} & \begin{tabular}[c]{@{}c@{}}81.75\\{[}78.75 - 84.72]\end{tabular} & \begin{tabular}[c]{@{}c@{}}72.60\\{[}69.29 - 75.67]\end{tabular} & \begin{tabular}[c]{@{}c@{}}0.4456\\{[}0.3798 - 0.5074]\end{tabular}  \\
        LSTE(ours) & \begin{tabular}[c]{@{}c@{}}\textbf{78.63}\\{[}76.23 - 81.00]\end{tabular} & \begin{tabular}[c]{@{}c@{}}\textbf{76.97}\\{[}74.77 - 79.14]~\end{tabular} & \begin{tabular}[c]{@{}c@{}}\textbf{0.9045}\\{[}0.8913 - 0.9173]\end{tabular} & \begin{tabular}[c]{@{}c@{}}\textbf{0.2856}\\{[}0.2518 - 0.3199]\end{tabular} & \begin{tabular}[c]{@{}c@{}}\textbf{87.90}\\{[}85.61 - 90.18]\end{tabular} & \begin{tabular}[c]{@{}c@{}}\textbf{78.85}\\{[}76.00 - 81.70]\end{tabular} & \begin{tabular}[c]{@{}c@{}}\textbf{0.5816}\\{[}0.5265 - 0.6365]\end{tabular}  \\
        \midrule
        w/o OrA           & \begin{tabular}[c]{@{}c@{}}73.70\\{[}71.16 - 76.20]\end{tabular} & \begin{tabular}[c]{@{}c@{}}72.93\\{[}70.62 - 75.21]\end{tabular}  & \begin{tabular}[c]{@{}c@{}}0.8867\\{[}0.8722 - 0.9007]\end{tabular} & \begin{tabular}[c]{@{}c@{}}0.3388\\{[}0.3025 - 0.3755]\end{tabular} & \begin{tabular}[c]{@{}c@{}}84.50\\{[}81.92 - 87.25]\end{tabular} & \begin{tabular}[c]{@{}c@{}}75.48\\{[}72.60 - 78.72]\end{tabular} & \begin{tabular}[c]{@{}c@{}}0.5042\\{[}0.4461 - 0.5694]\end{tabular} \\
        w/o Piecewise            & \begin{tabular}[c]{@{}c@{}}74.32\\{[}71.76 - 76.85]\end{tabular} & \begin{tabular}[c]{@{}c@{}}72.57\\{[}70.26 - 74.86]\end{tabular}  & \begin{tabular}[c]{@{}c@{}}0.8840\\{[}0.8687 - 0.8986]\end{tabular} & \begin{tabular}[c]{@{}c@{}}0.3480\\{[}0.3094 - 0.3871]\end{tabular} & \begin{tabular}[c]{@{}c@{}}83.33\\{[}80.60 - 86.23]\end{tabular} & \begin{tabular}[c]{@{}c@{}}76.92\\{[}74.10 - 80.03]\end{tabular} & \begin{tabular}[c]{@{}c@{}}0.5335\\{[}0.4764 - 0.5961]\end{tabular} \\
        w/o \(\mathcal{L}_{\text{PC}}\)           & \begin{tabular}[c]{@{}c@{}}73.90\\{[}71.32 - 76.45]\end{tabular} & \begin{tabular}[c]{@{}c@{}}72.36\\{[}70.03 - 74.65]\end{tabular}  & \begin{tabular}[c]{@{}c@{}}0.8825\\{[}0.8671 - 0.8973]\end{tabular} & \begin{tabular}[c]{@{}c@{}}0.3522\\{[}0.3134 - 0.3917]\end{tabular} & \begin{tabular}[c]{@{}c@{}}74.64\\{[}71.36 - 78.18]\end{tabular} & \begin{tabular}[c]{@{}c@{}}66.35\\{[}63.32 - 69.86]\end{tabular} & \begin{tabular}[c]{@{}c@{}}0.3354\\{[}0.2768 - 0.4036]\end{tabular}  \\
        \(\mathcal{L}_{\text{PC}}\) w/o \( w \)            & \begin{tabular}[c]{@{}c@{}}74.17\\{[}71.81 - 76.54]\end{tabular} & \begin{tabular}[c]{@{}c@{}}71.16\\{[}68.80 - 73.50]\end{tabular}  & \begin{tabular}[c]{@{}c@{}}0.8792\\{[}0.8640 - 0.8941]\end{tabular} & \begin{tabular}[c]{@{}c@{}}0.3600\\{[}0.3215 - 0.3986]\end{tabular} & \begin{tabular}[c]{@{}c@{}}74.76\\{[}71.21 - 77.94]\end{tabular} & \begin{tabular}[c]{@{}c@{}}66.35\\{[}63.09 - 69.75]\end{tabular} & \begin{tabular}[c]{@{}c@{}}0.4031\\{[}0.3608 - 0.4453]\end{tabular} \\
        \bottomrule
        \end{tabular}
        \raggedright{\scriptsize{$^1$The isolated backbone implementation of GLOMIA-Pro on ESCC. $^2$ The shared backbone implementation of GLOMIA-Pro on KOA.}}
    \label{tab5}
\end{table*}

\subsection{Comparison with Other Longitudinal Methods and generalize GLOMIA-Pro to Other Vision Backobones}
We perform a comprehensive evaluation of GLOMIA-Pro against seven state-of-the-art longitudinal analysis methods: LongFormer~\cite{chen2024longformer}, LOMIA-T~\cite{sun2024lomia}, TLSTM~\cite{yin2022predicting}, LP~\cite{ouyang2020longitudinal}, RP-Net~\cite{rosenberg2018prediction}, and DiT~\cite{tong2022dual}. These methods represent two dominant paradigms in longitudinal modeling: (1) longitudinal co-learning (LongFormer, LOMIA-T) and (2) longitudinal fusion (TLSTM, LP, RP-Net, DiT). Notably, our preliminary work LOMIA-T integrates both paradigms. Among these, RP-Net specializes in multi-task learning for simultaneous pCR prediction and tumor segmentation, while TLSTM incorporates a progression time matrix. These two methods were implemented exclusively on ESCC and KOA datasets, respectively. For our comparative analysis, we employ either the original implementations or meticulously reconstructed versions based on published methodologies. All methods are adapted to handle variable input dimensions, consistently achieving performance that was comparable to or surpassed the results reported in their respective original studies.

As shown in Table~\ref{tab1}, GLOMIA-Pro achieves superior performance across both datasets. For KOA severity prediction, GLOMIA-Pro attains a BAC of 78.63\%, demonstrating near-perfect agreement with radiologists (Cohen's Kappa = 0.90). It outperforms the second-best method LongFormer (76.44\%, \(p<0.001\)) by 2.86\% in BAC. In predicting pCR for ESCC patients, GLOMIA-Pro achieves an AUC of 87.90\% and an ACC of 78.85\%, improving upon the second-best method LOMIA-T (84.85\%, \(p<0.001\)) by 1.21\%. The integration of longitudinal co-learning, longitudinal fusion, and progression information as prior knowledge enables GLOMIA-Pro to effectively capture the temporal dynamics of longitudinal data, resulting in enhanced predictive performance.  

In binary classification of early KOA (grade 0 vs 2), GLOMIA-Pro achieves 87.32\% ACC, a 5.80\% improvement over existing diagnostic models~\cite{nasser2020discriminative} (82.53\%, Table~\ref{tab2}), demonstrating particular value for early intervention scenarios.

To validate the robustness of the framework, we implement it with four different cross-sectional backbones: ResNet50~\cite{he2016deep}, ResNet34 with CBAM~\cite{zhang2020attention}, DenseNet169~\cite{thomas2020automated}, and Swin-Transformer~\cite{liu2021swin} for KOA severity prediction. As presented in Table~\ref{tab3}, GLOMIA-Pro consistently enhances the performance of these models (\(p<0.001\)), improving BAC by 5.26\%, 6.26\%, 5.47\%, and 5.87\%, while reducing MAE by 22.67\%, 13.69\%, 16.57\%, and 20.97\%, respectively. Further validation on the ESCC dataset for pCR prediction demonstrates similar improvements. As shown in Table~\ref{tab4}, when ResNet50 or T2T-Transformer \cite{yuan2021tokens} is used as the backbone, GLOMIA-Pro boosts AUC by 14.24\% and 20.91\% and increases MCC by 43.17\% and 67.61\%, respectively. These results confirm the effectiveness and robustness of the GLOMIA-Pro.

\subsection{Ablation Study}  
To assess the contributions of key components in GLOMIA-Pro, we conduct ablation studies on the feature embedding module, spatiotemporal embedding, orthogonal attention mechanism, piecewise mechanism, and ordinal progression constraint (\(\mathcal{L}_{\text{PC}}\)). The effectiveness of the fusion strategy which integrates the progression representation ($\mathbf{P}$) with the current representation ($\mathbf{T}_\text{curr}$) has been demonstrated in our previous study \cite{sun2024lomia}.

\subsubsection{Task-specific weight sharing strategies for Feature Embedding} We evaluate different weight-sharing strategies (shared vs. isolated) for feature embedding module across both datasets, revealing dataset-specific optimal configurations shown in Table~\ref{tab5}. For KOA severity prediction, shared weights improve BAC by 3.57\% and reduce MAE by 13.35\% (both $p<0.001$). Conversely, pCR prediction in ESCC patients benefits from isolated weights, improving AUC and MCC by 3.22\% and 12.44\%, respectively (both $p<0.001$). This discrepancy arises from the nature of the datasets and the tasks at hand. The KOA dataset focuses relatively minor variations in joint structures. In this case, a shared embedding effectively captures these variations. Conversely, the ESCC dataset involves significant differences in images before and after treatment, with varying characteristics and importance at different time points. A separated embedding is more suitable for processing these distinct features.

\subsubsection{Effectiveness of spatiotemporal embedding (LSTE)} We compare our learnable spatiotemporal embedding (LSTE) with three alternatives: sinusoidal spatiotemporal embeddings (TPE) \cite{hu2023glim}, sinusoidal position embeddings combined with learnable temporal embeddings along feature dimension (TDPE) \cite{tong2022dual}, and learnable position embeddings combined with learnable temporal embeddings along image dimension (TIPE) \cite{sun2024lomia}. As shown in Table~\ref{tab5}, LSTE achieves 78.63\% BAC on KOA dataset and 87.90\% AUC ESCC datasets, outperforming those alternatives by 1.21\% BAC (76.62\%-77.69\%) and 5.80\% AUC (81.51\%-83.08\%). These results validate LSTE's enhanced capacity for modelling complex spatiotemporal relationships in longitudinal imaging data.

\subsubsection{Impact of Orthogonal Attention and Piecewise Mechanism} Ablation results confirm the critical role of orthogonal attention, as replacing it with standard attention leads to significant performance degradation (-6.69\% BAC on KOA, -4.02\% AUC on ESCC; both $p<0.001$). This might attribute to standard attention's tendency to overemphasize similar semantic features while suppressing clinically relevant temporal variations. Similarly, removing the piecewise mechanism results in significant declines in performance (-5.80\% BAC on KOA, -5.48\% AUC on ESCC; both $p<0.001$), underscoring its importance in preserving long-range dependencies essential for modeling disease progression. 

\subsubsection{Effectiveness of the Ordinal Progression Constraint}  
To evaluate the effectiveness of the ordinal progression constraint \(\mathcal{L}_{\text{PC}}\), we conduct ablation studies from two aspects: (i) removing \(\mathcal{L}_{\text{PC}}\) entirely and (ii) eliminating its ordinal properties. The first ablation results in substantial performance degradation (\(-6.41\%\) BAC on KOA, \(-17.77\%\) AUC on ESCC; both \(p<0.001\)). Notably, in the absence of \(\mathcal{L}_{\text{PC}}\) (73.90\% BAC on KOA and 74.64\% AUC on ESCC), the framework degenerates into a cross-sectional study with feature fusion, aligning with our findings (74.27\% BAC on KOA and 72.70\% AUC on ESCC, as shown in Table~\ref{tab3}, in row of baseline). The constraint \(\mathcal{L}_{\text{PC}}\) structures the latent representation space by clustering similar progression states while separating dissimilar ones, thereby enhancing the model's capability to preserve temporal dynamics. The second ablation, which removes the ordinal relationships, leads to similarly pronounced declines (\(-5.97\%\) BAC on KOA, \(-17.58\%\) AUC on ESCC; both \(p<0.001\)), underscoring the clinical reality that disease progression is a continuous process rather than discrete stages. By enforcing ordinal consistency, \(\mathcal{L}_{\text{PC}}\) directs the model's attention to subtle lesion changes while maintaining awareness of disease progression, thereby improving prediction performance for disease progression.

\section{Discussion and Conclusion}
In this study, we propose a Generalizable LOngitudinal Medical Imaging Analysis framework for predicting disease Progression (GLOMIA-Pro). GLOMIA-Pro leverages cross-sectional backbone networks to embed longitudinal imaging data into a representation space while employing piecewise orthogonal attention to model disease progression. An ordinal progression constraint is proposed to enforce ordinal consistency in disease stages, enhancing feature discrimination for longitudinal tasks. Additionally, progression aware fusion module integrates clinical and temporal priors to mitigate representation collapse, further improving model robustness. By effectively capturing the temporal dynamics of longitudinal data, the proposed framework GLOMIA-Pro achieves superior performance in predicting KOA severity and pCR outcomes in ESCC patients, outperforming seven state-of-the-art longitudinal methods across two clinical datasets and demonstrating strong robustness and generalizability.

Despite its superior performance, several limitations remain. First, while GLOMIA-Pro has been validated on both public and in-house clinical datasets, further evaluation across diverse clinical scenarios is necessary to confirm its generalizability. Second, although leveraging imaging from the most recent time point yields optimal predictive performance in two timepoints scenarios, incorporating multiple consecutive timepoint images could provide additional insights into disease progression. Third, recent advances in longitudinal pre-training \cite{ouyang2022disentangling, ouyang2022self} could further enhance GLOMIA-Pro by embedding temporal priors directly into the backbone network. Finally, while GLOMIA-Pro effectively captures complex temporal dependencies, its robustness to irregular follow-up intervals requires further investigation to ensure reliability in real-world clinical settings.


\bibliographystyle{IEEEtran}
\normalem
\bibliography{IEEEabrv, LOMIA}

\end{document}